\documentclass[%
 reprint,
 amsmath,amssymb,
 aps,
]{revtex4-2}

\usepackage{graphicx}
\usepackage{dcolumn}
\usepackage{bm}
\usepackage{braket}
\usepackage{tikz}
\usetikzlibrary{quantikz2}
\usepackage{physics}
\usepackage{appendix}
\usepackage{graphicx}
\usepackage{subcaption}

\begin{document}

\preprint{APS/123-QED}

\title{Shallow Quantum Scalar Products with Phase Information}

\author{Lila Cadi Tazi}
 \email{lc892@cam.ac.uk}
\affiliation{Yusuf Hamied Department of Chemistry, University of Cambridge, UK
}%

\author{David Muñoz Ramo}
\affiliation{Quantinuum, Terrington House, 13-15 Hills Road, Cambridge CB2 1NL, UK
}%

\author{Alex J. W. Thom}
\affiliation{Yusuf Hamied Department of Chemistry, University of Cambridge, UK \,
}%

\date{\today}

\begin{abstract}
The measurement of scalar products between two vectors is a common task in scientific computing and, by extension, in quantum computing. In this work, we introduce two alternative quantum circuits for computing scalar products with phase information, combining the structure of the {\sc swap} test, the vacuum test, and the Hadamard test. These novel frameworks, called the zero-control and one-control tests, present different trade-offs between circuit depth and qubit count for accessing the scalar product between two quantum states. We demonstrate that our approach significantly reduces the gate count for large numbers of qubits and decreases the scaling of quantum requirements compared to the Hadamard test.
\end{abstract}

\maketitle


\section{Introduction}

The measurement of scalar products between two vectors is a common task in scientific computing and, by extension, in quantum computing. In the quantum literature, several schemes already exist to perform this operation such as the {\sc swap} test measuring the squared scalar product or the Hadamard test. In this work, we introduce an alternative quantum circuit for computing scalar products, combining the structure of the {\sc swap} test and the Hadamard test. This novel framework is named the one-control test. It allows for the computation of scalar products with phase information and with fewer two-qubit gates than the Hadamard test. This is made possible by the introduction of classical information in the post-processing.
To explain our algorithm design, we first review the {\sc swap} test and the Hadamard test architectures.

We consider two arbitrary $n$-qubit quantum states $\ket{A}$ and $\ket{B}$. The target quantity is $\braket{B}{A}$.
We denote $U_A$ and $U_B$ the unitary gates that prepare $\ket{A}$ and $\ket{B}$, respectively.

\begin{equation}
  U_A \ket{0}^{\otimes n} = \ket{A} \quad\quad  U_B \ket{0}^{\otimes n} = \ket{B}
\end{equation}

\section{Existing Approaches}

\subsection{{\sc swap} Test}
\begin{figure}[h]
\centering
    \begin{quantikz}[]
        \lstick{$\ket{0}$} \qw & & \gate{H}& \ctrl{2} &  \gate{H} & \meter{} \\
        \lstick{$\ket{0}$}  & \qwbundle{n}&  \gate{U_A} & \swap{1} & & \\
        \lstick{$\ket{0}$}  & \qwbundle{n} & \gate{U_B} & \swap{-1} & & 
    \end{quantikz}
    \caption{{\sc swap} test circuit.}
    \label{fig:stest}
\end{figure}
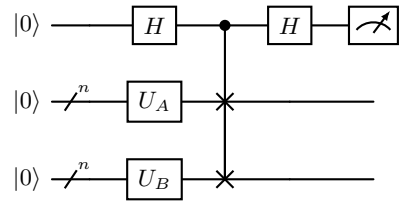

The {\sc swap} test is a widely used quantum algorithm that allows one to compute the squared overlap between two quantum states using the circuit in figure \ref{fig:stest} \cite{Buhrman2001}.

 Given two quantum registers with the prepared states $\ket{A}$ and $\ket{B}$, and an ancillary qubit initialised to $\ket{0}$, the {\sc swap} test is constructed by applying a Hadamard gate on the ancillary, followed by a control {\sc swap}  ({\sc cswap}) operation and another Hadamard gate on the ancillary.
$Z$ measurements of the ancillary qubit return $\ket{0}$ with probability $p(0) = \frac{1}{2} + \frac{1}{2}|\braket{B}{A}|^2$ and $\ket{1}$ with probability $p(1) = \frac{1}{2} - \frac{1}{2}|\braket{B}{A}|^2$.
The squared scalar product can then be accessed classically from the results of the quantum measurement :

\begin{equation}
    |\braket{B}{A}|^2 = p(0) - p(1).
\end{equation}

This algorithm requires relatively few quantum gates. It can be implemented after the state preparation procedures without affecting the $U_A$ and $U_B$ circuits. The most expensive operation is the {\sc cswap}, which can be decomposed into 8 {\sc cnot} gates per qubit across the swapped registers, that is, $8n$ {\sc cnot} gates here.
However, the measured quantity is the squared scalar product, so by using this method we lose all phase information of the scalar product. To our knowledge, there is no obvious way to retrieve the phase of the scalar product.

\subsection{Vacuum Test}

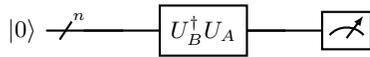
\begin{figure}[ht]
    \centering
    \begin{quantikz}[]
        \lstick{$\ket{0}$}  & \qwbundle{n} &  \qw  & \gate{U_{B}^\dagger U_A}  &  \qw     & \meter{}
    \end{quantikz}
    \caption{Vacuum test circuit for computing $|\braket{B}{A}|^2$.}
    \label{fig:Vacuum_Test}
\end{figure}

The vacuum test is an alternative to the swap test to compute the squared scalar product between two states \cite{Lee2018}. The circuit structure is shown in figure \ref{fig:Vacuum_Test}, where $U_B^\dagger U_A$ is applied to the $\ket{0}$ state. By measuring the register in the computational basis, the expectation value $|\bra{0}U_B^\dagger U_A\ket{0}|^2 = |\braket{B}{A}|^2$ can be estimated, which corresponds to the probability of measuring the all-zero state.

Compared to the {\sc swap} test, this algorithm requires less qubits but increases the circuit depth. A comparative performance of both algorithms was investigated in \textcite{Tudorovskaya2024,Feniou}.

\subsection{Hadamard test}

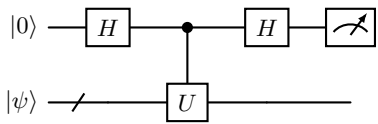
\begin{figure}[h]
\centering
    \begin{quantikz}
   \lstick{$\ket{0}$} \qw & \gate{H}& \ctrl{1}&  \gate{H} & \meter{} \\
    \lstick{$\ket{\psi}$}  &   \qwbundle{} & \gate{U} & & \\
    \end{quantikz}
\caption{Hadamard test circuit}
\label{fig:htest}
\end{figure}

The Hadamard test is a quantum algorithm that allows one to measure the expectation value of a unitary operator \cite{Cleve1998}. Given a unitary operator $U$ and a state $\ket{\psi}$, the expectation value $\bra{\psi}U\ket{\psi}$ can be measured by constructing the circuit in figure \ref{fig:htest}. An ancillary qubit initialised to $\ket{0}$ is added to the circuit containing the state $\ket{\psi}$. The controlled version of the unitary operator $U$ is added between two Hadamard gates on the ancillary.

When measuring the ancilla qubit in the $Z$ basis, the probabilities of outcomes are :
\begin{equation}
    \begin{aligned}
         & p(0) = \frac{1}{4} \bra{\psi}(I+U^\dagger)(I+U)\ket{\psi}, \\
         & p(1) = \frac{1}{4} \bra{\psi}(I-U^\dagger)(I-U)\ket{\psi}.
    \end{aligned}
\end{equation}
Thus,
\begin{equation}
    p(0)-p(1) = \frac{1}{2} \bra{\psi}(U^\dagger+U)\ket{\psi} = \Re(\bra{\psi}U\ket{\psi}),
\end{equation}

\noindent obtaining the real part of the expectation value $\Re(\langle U \rangle)$.
Likewise, the imaginary part can be obtained by replacing the first $H$ gate with $HS^\dagger$ creating the state $\frac{1}{\sqrt{2}}(\ket{0}-i\ket{1})$ in the ancillary qubit.

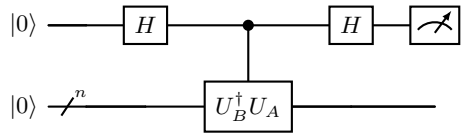
\begin{figure}[ht]
    \centering
    \begin{quantikz}[]
        \lstick{$\ket{0}$}  &  \qw      & \gate{H}  & \ctrl{1}  & \gate{H}  & \meter{} \\
        \lstick{$\ket{0}$}  & \qwbundle{n} &  \qw  & \gate{U_{B}^\dagger U_A}  &  \qw     & \qw
    \end{quantikz}
    \caption{Hadamard test circuit for computing $\Re(\braket{B}{A})$.}
    \label{fig:ScalarProd_H_Test}
\end{figure}

The Hadamard test architecture can be used to measure scalar products by taking inspiration from the vacuum test architecture and computing the expectation value of $U_{B}^\dagger U_A$ on state $\ket{0}^{\otimes n}$, as shown in figure \ref{fig:ScalarProd_H_Test}.

\noindent Upon measurement, we obtain 
\begin{equation}
    p(0)-p(1) = \Re(\bra{0}U_{B}^\dagger U_A\ket{0}) = \Re(\braket{B}{A})
\end{equation}
and $\Im(\braket{B}{A})$ is obtained by first applying $HS^\dagger$ to the ancillary qubit.
With this circuit, one can measure $\braket{B}{A}$ with phase information. However, it requires to compute $U_B^\dagger$ the conjugate transpose of $U_B$ and implement the controlled version of $U_A$ and $U_B^\dagger$.

A naive implementation of the controlled versions of $U_A$ and $U_B^\dagger$
can increase the number of two-qubit gates by up to four times compared to the original unitaries \cite{mottonen}. Specifically, in terms of {\sc cnot} gates, this overhead can be up to six times the original count \cite{Krol2022}.
The circuit depth is also substantially impacted since the ancillary qubit is always acted upon. 
The controlled state preparation operation can be implemented more efficiently using ancillary qubits \cite{Yuan2023}.
The overhead in two-qubit gate count and circuit depth can be prohibitive.

\subsection{Overlap Estimation Algorithm}
The Overlap Estimation Algorithm was proposed in \textcite{Policharla2021}. It relies on preparing the state $\frac{1}{\sqrt{2}}(\ket{0}\ket{A}+\ket{1}\ket{B})$ and then measuring the ancillary qubit similarly to a Hadamard test. It enables the computation of the phase of $\braket{B}{A}$. This method requires the implementation of two controlled unitaries, with $U_A$ controlled on $\ket{0}$ and $U_B$ controlled on $\ket{1}$. It has a similar cost to the Hadamard test.

\section{New Frameworks Reducing Control Overhead}

\subsection{One-control test}

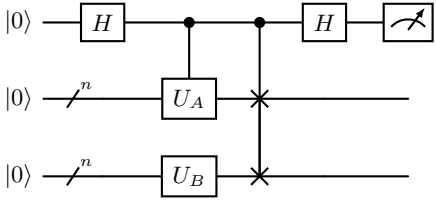
\begin{figure}[h]
\centering
    \begin{quantikz}
   \lstick{$\ket{0}$} \qw & \gate{H}& \ctrl{1} & \ctrl{2} &  \gate{H} & \meter{} \\
    \lstick{$\ket{0}$}  & \qwbundle{n} & \gate{U_A} & \swap{1} & & \\
    \lstick{$\ket{0}$}  & \qwbundle{n} & \gate{U_B} & \swap{-1} & & 
    \end{quantikz}
\caption{One-control test circuit for $\Re(\braket{B}{A})$.}
\label{fig:one-control}
\end{figure}

We propose an alternative circuit for computing the scalar product of two quantum states, called the one-control test. It takes inspiration from the {\sc swap} test and Hadamard test circuits but requires controlling only one of the preparation unitaries before applying a {\sc cswap}. This method avoids the presence of a squared term when measuring the ancilla qubit, allowing us to compute the phase of the scalar product. The circuit structure is shown in figure \ref{fig:one-control}.

To extract the scalar product $\braket{B}{A}$, we need the coefficient $b_0$ corresponding to $\braket{0...00}{B}$. In some cases, this coefficient can be known classically, or we can estimate it by measuring the $B$ state, or by performing another one-control test procedure with $\ket{A}=\ket{0}^{\otimes n}$.

When measuring the ancilla qubit the probabilities of outcomes are 
\begin{equation}
    \begin{aligned}
         & p(0) = \frac{1}{4} (\bra{0} \otimes \bra{B} + \bra{B} \otimes \bra{A}) (\ket{0} \otimes \ket{B} + \ket{B} \otimes \ket{A}) \\
        &=  \frac{1}{2} +  \frac{1}{2} \Re(\braket{B}{A} \braket{B}{0}) =  \frac{1}{2} +  \frac{1}{2} \Re(\braket{B}{A} b_0) \\
         & p(1) = \frac{1}{4}  (\bra{0} \otimes \bra{B} - \bra{B} \otimes \bra{A}) (\ket{0} \otimes \ket{B} - \ket{B} \otimes \ket{A}) \\
        & =  \frac{1}{2} - \frac{1}{2} \Re(\braket{B}{A} \braket{B}{0}) =  \frac{1}{2} - \frac{1}{2} \Re(\braket{B}{A} b_0),
    \end{aligned}
\end{equation}
and
\begin{equation}
    R = p_{\Re}(0)-p_{\Re}(1) = \Re(\braket{B}{A} b_0).
\end{equation}

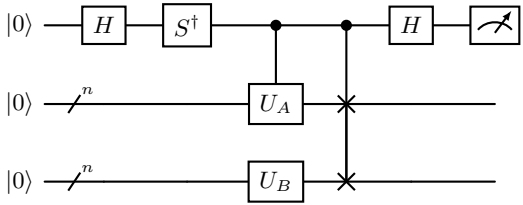
\begin{figure}[h]
\centering
    \begin{quantikz}
   \lstick{$\ket{0}$} \qw & \gate{H}& \gate{S^\dagger}& \ctrl{1} & \ctrl{2} &  \gate{H} & \meter{} \\
    \lstick{$\ket{0}$}  &\qwbundle{n} && \gate{U_A} & \swap{1} & & \\
    \lstick{$\ket{0}$}  & \qwbundle{n} && \gate{U_B} & \swap{-1} & & 
    \end{quantikz}
\caption{Imaginary part one-control test circuit}
\label{fig:Im one-control}
\end{figure}

We can compute the imaginary part by adding a $S^\dagger$ gate after the first Hadamard, similarly to the Hadamard test, with the circuit in figure \ref{fig:Im one-control}. We then obtain
\begin{equation}
      I = p_{\Im}(0)-p_{\Im}(1) = \Im(\braket{B}{A} b_0).
\end{equation}

Decomposing the real and imaginary parts of $b_0$
\begin{equation}
    b_0 = c + i d,
\end{equation}

we obtain the scalar product with equation \ref{eq:braketOneControl}.

\begin{equation}\label{eq:braketOneControl}
     \braket{B}{A} = \frac{cR + dI}{|b_0|^2}  + i \frac{cI - dR}{|b_0|^2}
\end{equation}

\noindent This algorithm only works for nonzero $b_0$ coefficient. If $|\braket{0...00}{B}|=0$, the projection should be performed on another non-zero basis state, as described in section \ref{sec:arbitraryb0}.

This algorithm requires the injection of classical information $b_0$ to compute the scalar product. Compared to the Hadamard test, it has only one controlled preparation gate, thus significantly reducing the two-qubit gate count. 

This circuit framework can be particularly useful when the $U_A$ preparation gate has a simple form, for example if it creates a product state, as the number of two-qubit gates in the controlled $U_A$ gate can be sufficiently small in this case.

\subsection{Zero-control test}

\begin{figure}[h]
\centering
    \begin{quantikz}
   \lstick{$\ket{0}$} \qw & \gate{H}& & \ctrl{1} & \ctrl[open]{2} &  \gate{H} & \meter{} \\
   \lstick{$\ket{0}$}  & \qwbundle{n} & & \swap{1} & \swap{2} & & \\
    \lstick{$\ket{0}$}  & \qwbundle{n} & \gate{U_A} & \swap{-1} & & & \\
    \lstick{$\ket{0}$}  & \qwbundle{n} & \gate{U_B} &  & \swap{-2} &  &
    \end{quantikz}
\caption{Zero-control test circuit for $\Re(\braket{B}{A})$.}
\label{fig:zero-control}
\end{figure}
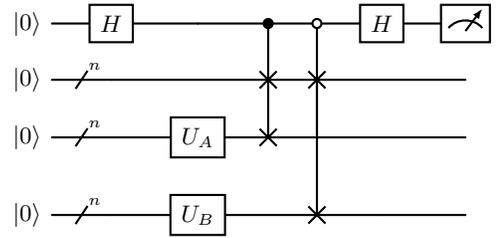

Applying the same concept once more and drawing from the overlap estimation algorithm \cite{Policharla2021}, we can further reduce the need for controlled unitaries. This innovative framework, called the zero-control test, is illustrated in figure \ref{fig:zero-control}.

Here, the $a_0$ coefficient corresponding to $\braket{0..00}{A}$ is also required to obtain the scalar product.
\begin{equation}
    R =  p_{\Re}(0)-p_{\Re}(1) = \Re(\braket{B}{A} a_0 b_0)
\end{equation}
The imaginary part can be computed in the same way as the one-control test, by adding a $S^\dagger$ gate to the ancillary qubit. Upon measurement we obtain :
\begin{equation}
    I = p_{\Im}(0)-p_{\Im}(1) = \Im(\braket{B}{A} a_0 b_0).
\end{equation}

Decomposing the real and imaginary parts of $a_0$ and $b_0$
    
\begin{equation}
\begin{aligned}
    &a_0 = e + i f \\
    &b_0 = c + i d, 
\end{aligned}
\end{equation}

we define : 

\begin{equation}
\begin{aligned}
    &g = ec - fd \\
    &h = ed + fc.
\end{aligned}
\end{equation}

The target scalar product is calculated with equation \ref{eq:zero-control_SP}.

\begin{equation}\label{eq:zero-control_SP}
     \braket{B}{A} = \frac{gR + hI}{|a_0|^2|b_0|^2}  + i \frac{gI - hR}{|a_0|^2|b_0|^2}
\end{equation}

This circuit uses $3n+1$ qubits in total and includes two controlled {\sc swap} gates, with one controlled on the $\ket{0}$ state of the auxiliary qubit. The $\ket{0}$-control can simply be implemented by adding one {\sc x} gate on each side of the {\sc cswap} \cite{nielsen2010}.
When both $a_0$ and $b_0$ are classically known, this method results in the circuit having the least depth. It only introduces the cost of two {\sc cswap}, that is $16n$ {\sc cnot} gates, but it uses $3n+1$ qubits which can be a limiting factor for quantum hardware implementations.

\subsection{Projecting on an arbitrary state}\label{sec:arbitraryb0}

We can choose an arbitrary state to project the $B$ amplitudes on, by adding some 0-{\sc cnot} gates ({\sc x} if the control state is $0$) before the state preparation.

\begin{figure}[h]
\centering
    \begin{quantikz}[row sep=0.1cm]
   \lstick{$\ket{0}$} \qw & \gate{H}& \octrl{3} & \ctrl{1} & \ctrl{2} &  \gate{H} & \meter{} \\
    \lstick{$\ket{0}$} &  & \targ{}  & \gate[3]{U_A} & \gate[4]{\textrm{{\sc swap}}} & & \\
    \lstick{$\ket{0}$} & \qwbundle{n-2} &   &  &  & & \\
    \lstick{$\ket{0}$} &  & \targ{}  &  &  & & \\
    \lstick{$\ket{0}$} & \qwbundle{n} &  & \gate{U_B} &  & & 
    \end{quantikz}
\caption{Example of one-control test circuit for projection on $\ket{10..01}$}
\label{fig:one-control-arbitrary-projection}
\end{figure}
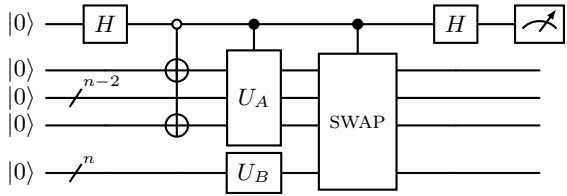

To project the state $B$ to an arbitrary bit string state in the computational basis, the 0-{\sc cnot} gates should be applied to all qubits equal to one in the target bit string.

\subsection{Numerical Validation}

To balance the trade-off between circuit depth and gate count, we focused on studying the performances of the one-control test.

The primary benefit of the one-control test lies in the requirement to control just a single preparation unitary. This becomes especially beneficial if $U_A$ can be represented by a shallow quantum circuit. A practical use case for the one-control or zero-control tests is the assessment of a phase-based state quality metric, since phaseless measurements such as fidelity are not always sufficient \cite{Blunt2023, Benedetti2021}.

To demonstrate our algorithm, we applied the method to the computation of scalar products $\braket{B}{A}$ in the case where $\ket{A}$ is a separable state and $\ket{B}$ is a dense quantum state with random amplitudes. The preparation circuits were designed using the quantum Shannon decomposition scheme \cite{shende_synthesis_2006}.

\begin{figure}[h]
\centering
   $U_A$ =  \begin{quantikz}[row sep=0.2cm]
    \lstick[3]{$p$} & \qwbundle{n/p} & \gate{U_a}&   \\
     & \qwbundle{n/p} & \gate{...}  &  \\
     & \qwbundle{n/p} & \gate{U_a} &  \\
    \end{quantikz}
\caption{State preparation circuit used for the separable state. The $n$-qubits $U_a$ unitary is repeated $p$ times in parallel.}
\label{fig:uA_sep}
\end{figure}
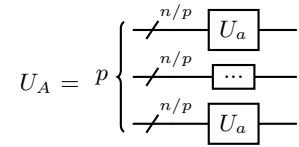

We construct $\ket{A}$ on $n$ qubits by applying a $\frac{n}{p}$-qubits preparation gate $p$ times in parallel as shown in figure \ref{fig:uA_sep}. 

This creates a product state in the form 
\begin{equation}
    \ket{A} = U_A \ket{0} =  \bigotimes^p U_a \ket{0} = \bigotimes^p \ket{a}.
\end{equation}

\begin{figure}[ht]
    \centering
        \centering
        \includegraphics[width=0.5\textwidth]{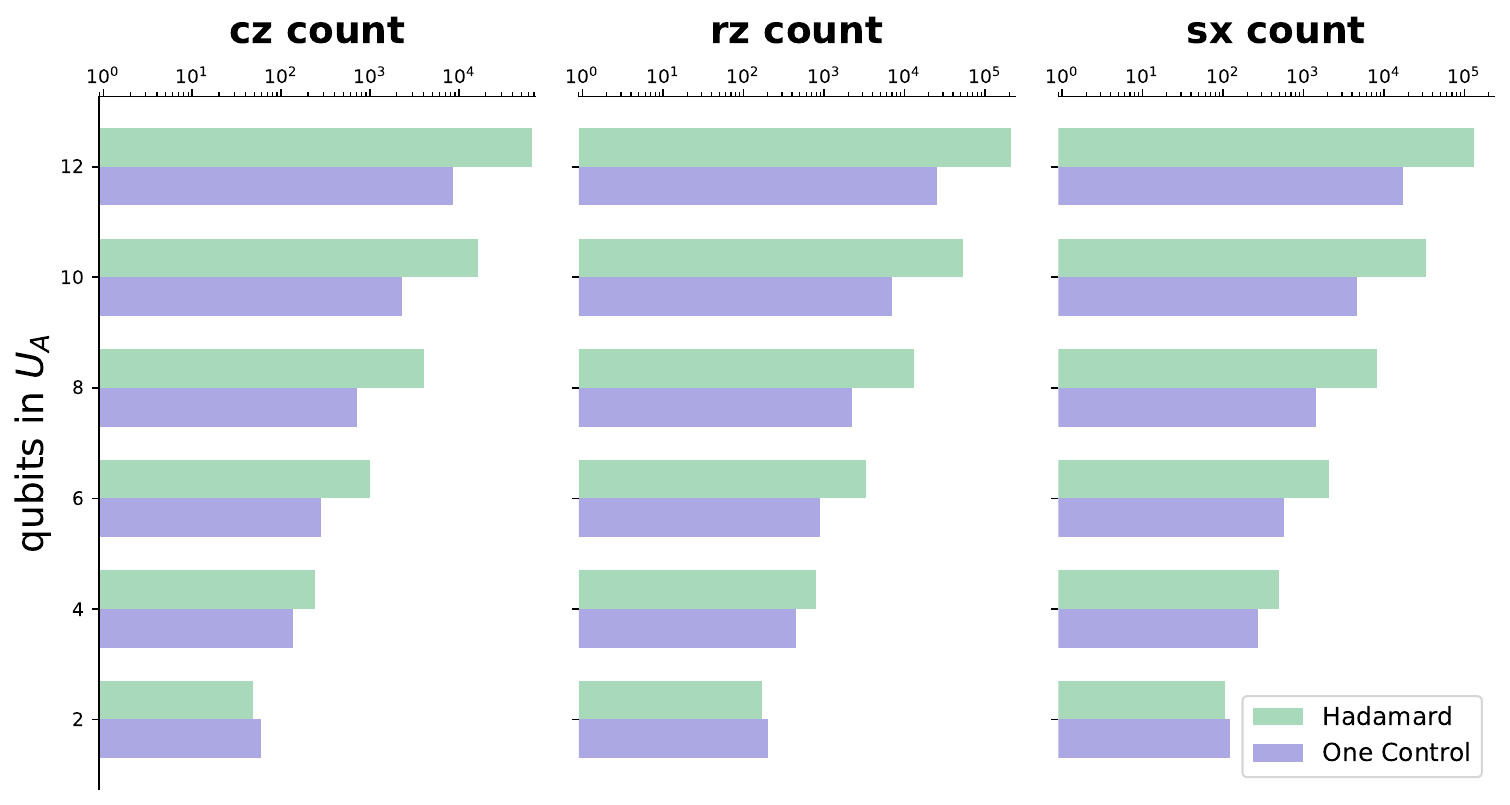}
        \caption{Gate count comparison between the Hadamard test and the one-control test. A constant cost of two additional {\sc x} gates for the one-control test is not reported in the figure.}
        \label{fig:gates_counts}
\end{figure}

\begin{figure}[ht]
        \centering
        \includegraphics[width=0.5\textwidth]{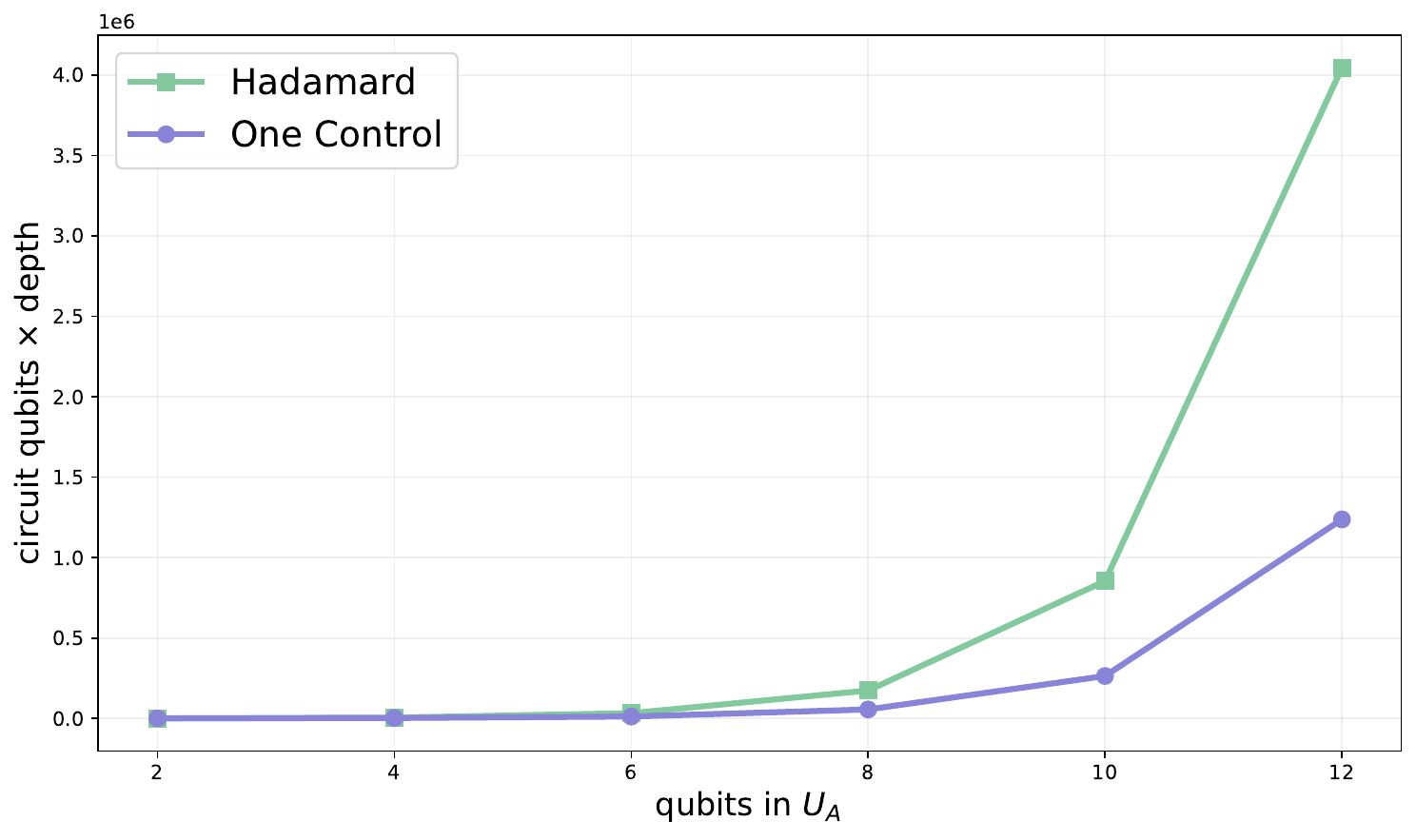}
        \caption{Quantum resource comparison between the Hadamard test and the one-control test. The reported numbers are the total number of qubits times the circuit depth.}
        \label{fig:resource}
\end{figure}

In figure \ref{fig:gates_counts}, we report the number of operators in the one-control and Hadamard test circuits.
The circuits were compiled to match the gate set available on IBM Heron hardware, that is \{{\sc cz, rz, sx, x}\}.
The numbers reported are the sum of all gate counts needed to compute the real and imaginary part of the scalar product. For more than $n=4$ qubits in $U_A$ the one-control test presents a reduced gate count compared to the Hadamard test. The reduction of the two-qubit gate count can be a major advantage for near-term algorithms, as this is often the limiting factor for practical quantum experiments.

In comparison to the Hadamard test, our method doubles the number of qubits in the main register. We examine the product of circuit depth and total qubit count, $d \times q$, as an indicator of the quantum resources needed for both circuits. The values are presented in figure \ref{fig:resource}.
For a small number of qubits, the Hadamard test circuit requires fewer gates as a result of the overhead of the {\sc cswap} gate in the one-control test. However, our method shows better scaling in circuit depth, making it more advantageous for larger qubit counts. 
To select the most suitable method, one can estimate the gate count in the controlled $U_B^\dagger$ unitary and compare to the $8n$ overhead introduced by the one-control test.

\section{Conclusion}
The zero-control and one-control tests can be alternative approaches to the well-known Hadamard test to compute scalar products with phase information. We have demonstrated that our approach significantly reduces the gate count for large numbers of qubits and reduces the scaling of quantum requirements compared to the Hadamard test.
 Our algorithm includes some classical information $a_0$ and $b_0$ in the post-processing step. If these coefficients are not known a priori, an additional overhead is required to measure them. Since the one-control test still requires the controlled version of one of the preparation unitaries, it shows a greater advantage when $U_A$ is a shallow circuit.
The zero-control test allows one to compute scalar products without having to control the preparation unitaries, but with a larger overhead in terms of number of qubits and with the requirement to access two classical coefficients $a_0$ and $b_0$.
Both approaches offer a trade-off between qubit and gate counts, where the Hadamard test requires $n+1$ qubits and two controlled unitaries, the one-control test needs $2n+1$ qubits and only one controlled unitary plus a {\sc cswap} gate, and finally the zero-control test requires $3n+1$ qubits and two {\sc cswap} gates but no control on the preparation unitaries.

\section*{Acknowledgements}

LCT gratefully acknowledges Quantinuum for funding this work. The authors thank Michelle Sze and Etienne Granet for their thorough review of the manuscript and valuable suggestions. We also thank David Zsolt Manrique and Cristina Cirstoiu for their insightful comments that enriched this study.

\bibliography{main}

\end{document}